\documentclass[reprint,superscriptaddress,amsmath,amssymb,aip,notitlepage,floatfix]{revtex4-1}
\usepackage{ulem} 
\usepackage[svgnames]{xcolor}
\usepackage{graphicx}

\usepackage{amsmath}

\usepackage[version=4]{mhchem}
\usepackage{siunitx}
\usepackage{miller}

\definecolor{ocre}{RGB}{243,102,25} 
\definecolor{verydarkred}{RGB}{150,0,0} 

\newcommand{\EuPdII}{EuPd$_2$}

\makeatletter
\def\@email#1#2{%
 \endgroup
 \patchcmd{\titleblock@produce}
  {\frontmatter@RRAPformat}
  {\frontmatter@RRAPformat{\produce@RRAP{*#1\href{mailto:#2}{#2}}}\frontmatter@RRAPformat}
  {}{}
}%
\makeatother

\begin{document}


\title{Magnetic properties of multi-domain epitaxial \ce{EuPd2} thin films}


\author{Alfons G. Schuck}
\affiliation{Institute of Physics, Goethe University Frankfurt, Max-von-Laue Street 1, Frankfurt am Main 60438, Germany}
\author{Sebastian K\"olsch}
\affiliation{Institute of Physics, Goethe University Frankfurt, Max-von-Laue Street 1, Frankfurt am Main 60438, Germany}
\author{Adrian Valadkhani}
\affiliation{Institute for Theoretical Physics, Goethe University Frankfurt, Max-von-Laue Street 1, Frankfurt am Main 60438, Germany}
\author{Igor I. Mazin}
\affiliation{Department of Physics and Astronomy, George Mason University, Fairfax, VA 22030}
\affiliation{Quantum Science and Engineering Center, George Mason University, Fairfax, VA 22030}
\author{Roser Valent\'i}
\affiliation{Institute for Theoretical Physics, Goethe University Frankfurt, Max-von-Laue Street 1, Frankfurt am Main 60438, Germany}
\author{Michael Huth}
\email[]{michael.huth@physik.uni-frankfurt.de}
\affiliation{Institute of Physics, Goethe University Frankfurt, Max-von-Laue Street 1, Frankfurt am Main 60438, Germany}

\date{\today}

\begin{abstract}
Europium intermetallic compounds show a variety of different ground states and anomalous physical properties due to the interactions between the localized 4f electrons and the delocalized electronic states. Europium is also the most reactive of the rare earth metals which might be the reason why very few works are concerned with the properties of Eu-based thin films. Here we address the low-temperature magnetic properties of ferromagnetic \ce{EuPd2} thin films prepared by molecular beam epitaxy. The epitaxial (111)-oriented thin films grow on MgO \hkl(100) with eight different domain orientations. We analyze the low-temperature magnetic hysteresis behavior by means of micromagnetic simulations taking the multi-domain morphology explicitly into account and quantify the magnetic crystal anisotropy contribution. By \textit{ab initio} calculations we trace back the microscopic origin of the magnetic anisotropy to thin film-induced uniform biaxial strain.
\end{abstract}


\maketitle
%
%
\section{Introduction}
Intermetallic Eu-compounds are an interesting class of materials due to the sensitivity of the Eu valence state 
to the available volume at the respective lattice sites. Eu can have two possible valence states; magnetic divalent \ce{Eu^{2+}} (4f$^7$) \ce{Eu(II)}
with a total angular momentum of $J=7/2$ and non-magnetic trivalent \ce{Eu^{3+}} (4f$^6$) Eu(III) with $J=0$. 
As a function of temperature, strain, or pressure, a number of Eu-based compounds undergo a valence crossover accompanied by significant volume changes due to the strong coupling of lattice and spin degrees of freedom \cite{Onuki2017, Song2023}. Here, thin-film induced effects, such as epitaxial clamping, can lead to pronounced changes of the materials' properties if compared to bulk crystals, as was recently shown for \ce{EuPd2Si2}, which exhibits a temperature-driven Eu valence transition in bulk crystals that is suppressed in epitaxial thin films due to clamping;  thus stabilized Eu(II) state, instead, leads to a magnetic phase transition \cite{Koelsch2022}. Also for Eu-compounds that show magnetic order in bulk form, thin-film induced changes of the magnetic state can be expected but have not been studied in detail so far. Here we study the ferromagnetic compound \ce{EuPd2}, which we have been able to grow epitaxially by molecular beam epitaxy.

From the about 200 known binary Laves phases \ce{AB2} with rare earth elements (A) and partner metals (B) from the group 7-10, \ce{EuPd2}, which forms in the cubic C15-type (space group Fd\=3m) is special, because Pd does not form a Laves phase with any other rare earth element \cite{Gschneidner2006}. Bulk EuPd$_2$ is a ferromagnet with the Curie temperature $T_C=  80$ K and a low-temperature saturation magnetization $M_s=1.19\times 10^6$\,A/m (i.\,e.\ $7.49\,\mu_B$/Eu(II)) that is solely based on the Eu(II) spin ($J=S=7/2$) moment; see, e.\,g.\ \cite{Nakamura2016, Gschneidner2006, Harris1972}.

\begin{figure}[ht]
\centering
\includegraphics[width=\linewidth]{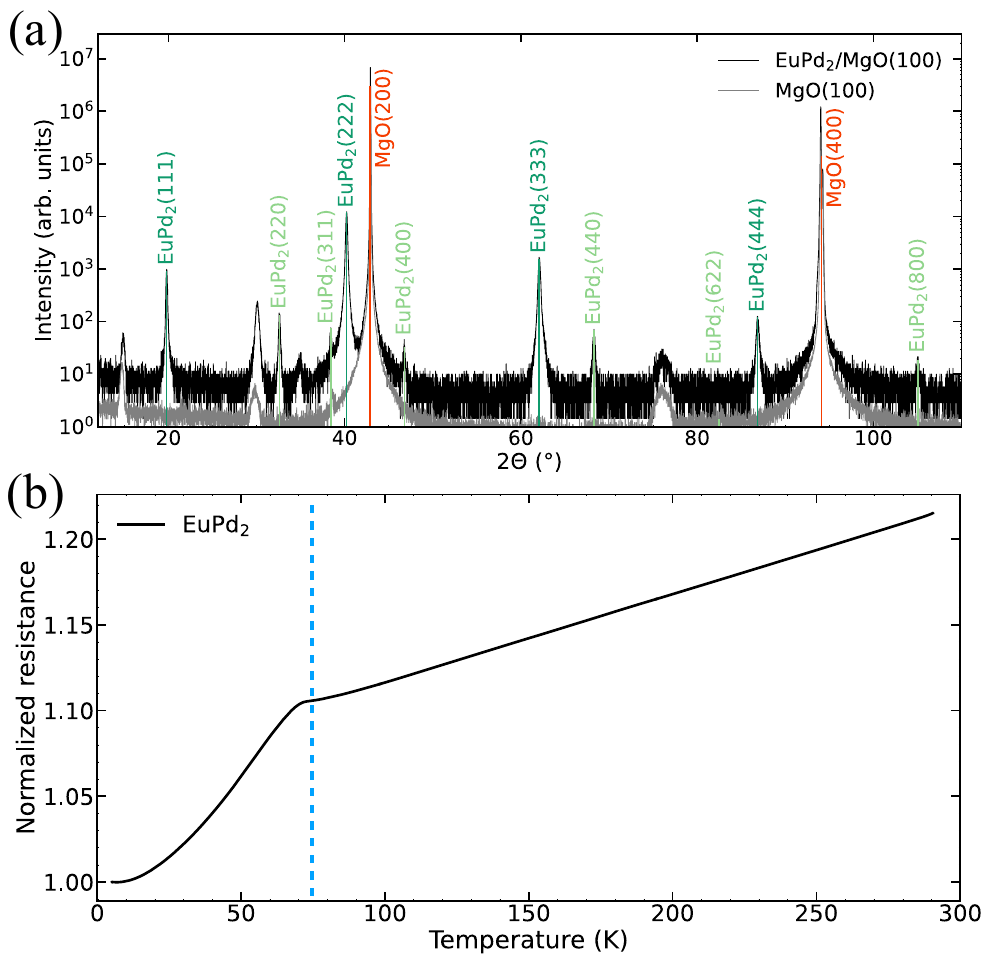}
\caption{(a) Symmetric Bragg scan (longitudinal X-ray scan) of \ce{EuPd2} thin film grown on \ce{MgO}\hkl(100) showing the preferred \hkl(111) growth orientation. (b) Temperature-dependent resistance data of the \ce{EuPd2} thin film normalized to the low-temperature residual resistance value. The onset of ferromagnetic order is indicated as the upturn before the kink in the resistance.}
\label{fig1}
\end{figure}

In a recent publication we presented results on the epitaxial thin film growth of the binary Eu-compounds \ce{EuPd3} and \ce{EuPd2} onto \ce{MgO}\hkl(100) substrates by means of molecular beam epitaxy (MBE) \cite{Koelsch2023}. In that work we focused on the growth preferences and epitaxial relationships, as well as on the structural and temperature-dependent magnetotransport properties of the thin films. As one of the main results, we found a pronounced \hkl(111) growth preference of \ce{EuPd2} although this orientation with sixfold symmetry does not match with the fourfold symmetry of the surface layer of cubic \ce{MgO}\hkl(100); see Fig.\,\ref{fig1}(a). From temperature-dependent resistivity data we could confirm a phase transition at about \qty{72}{K} (see Fig.\,\ref{fig1}(b)), which we attributed to the onset of ferromagnetic order. This was further corroborated by Hall effect and magnetoresistance measurements \cite{Koelsch2023}. An important consequence of the combined effects of lattice symmetry differences at the substrate--thin-film interface and weak thin-film substrate interaction, we observed an island growth mode with multiple growth domains; see below in Fig.\,\ref{fig2}(a,b).

In this work we focus on the properties of the ferromagnetic state of the \ce{EuPd2} thin films on \ce{MgO} at low temperatures taking this multi growth domain structure and morphology explicitly into account. We do this by comparing micromagnetic simulation results to magnetic hysteresis measurements performed at low temperature. The experimentally observed coercive field in the mT-range for magnetic field aligned in the thin film plane indicates a magnetic crystal anisotropy which is \textit{a priori} not expected for \ce{EuPd2}. As Eu is divalent in \ce{EuPd2} and has a closed f-shell in the spin-majority channel, it would have zero orbital moment if considered as a free \ce{Eu^{2+}} ion. This \ce{Eu} valence furthermore implies that Pd accepts one extra electron per ion becoming formally iso-electronic to Ag. As a free \ce{Pd-} ion, again the filled d-shell would have no orbital moment. As we show, the observed magnetic anisotropy is a consequence of Eu and Pd ions being not free, and their combined electronic bands are still affected by spin-orbit coupling, the consequences of which are discussed in conjunction with clamping-induced uniform biaxial strain forming as the samples are cooled to cryogenic temperatures.
%
%
\section{Methods}
%
%
\subsection{Thin film growth}
\ce{EuPd2} thin film growth was accomplished by MBE in a vacuum chamber with a base pressure well below \qty{1e-10}{mbar}, which is solely dedicated to the deposition of intermetallic compounds.
Prior to co-deposition of \ce{Eu} and \ce{Pd}, the \ce{MgO}\hkl(100) substrate was annealed under ultra high vacuum at an elevated temperature of \qty{1000}{\degreeCelsius} for one hour in order to improve surface cleanliness. Thin film growth was started after reaching the required stable sample temperature of \qty{450}{\degreeCelsius} at an average growth rate of about \qty{0.5}{nm/min}. To this end, the Eu and Pd flux rates were adjusted using a quartz crystal microbalance to yield the required stoichiometry. According to the results of the growth of epitaxial thin films of \ce{EuPd}{3} on \ce{MgO}\hkl(100) substrates, Eu desorption due to the high substrate temperature plays a major role during preparation of Eu-based films, see \cite{Koelsch2023}.
This so called MBE- or Eu-distillation effect also occurs during the formation of stoichiometric \ce{EuO} thin films for sufficiently high growth temperatures and with a chemically inert substrate at a low oxygen partial pressure \cite{Sutarto2009}. Subsequent to film preparation and cooling the sample down to room temperature, a protective amorphous silicon capping layer with a thickness of about \qty{5}{nm} was deposited to prevent surface oxidation. The structural properties of the film were investigated by X-ray diffraction using a Bruker D8 Discover high resolution diffractometer, employing monochromatized Cu$_{K,\,\alpha}$ radiation under ambient conditions in symmetric and asymmetric reflection geometry. As is evident from the longitudinal X-ray scan in Fig.\,\ref{fig1}\,(a), mainly \hkl(111)-oriented out-of-plane growth occurs.
Close inspection of the region around the \ce{EuPd2}\hkl(111) reflex reveals some Laue oscillations which are indicative of high structural order with a coherence length of about \qty{20}{nm}. Additionally, $\phi$-rotational scans in asymmetric geometry (see Fig.\,\ref{fig2}\,(b) confirm the existence of eight different in-plane growth domains, leading to the epitaxial relationship of:

\begin{equation*}
	\begin{split}
&\ce{MgO}\hkl{100} \parallel \ce{EuPd2}\hkl{111} \\
\mathrm{and}\, &\ce{MgO}\hkl<110> \parallel \ce{EuPd2}\hkl<10-1> \\
\mathrm{or}\, &\ce{MgO}\hkl<100> \parallel \ce{EuPd2}\hkl<10-1> 
\end{split}
\end{equation*}

Further analysis of the film's morphology was done by scanning electron microscopy (SEM) performed in an FEI Nova NanoLab 600 dual beam electron/ion microscope. According to the SEM investigation (see Fig.\,\ref{fig2}\,a) well defined faceted islands on a lateral length scale of about \qty{100}{nm} develop, thus corroborating the multi-domain results from the asymmetric X-ray scans. Further details regarding the microstructure and the magnetotransport properties can be found in Ref. \cite{Koelsch2023}.
%
%
\subsection{Magnetic measurements}
Magnetization measurements at cryogenic temperatures were performed using a vibrating sample magnetometer (VSM), custom-designed as retrofit for a \ce{^4He} variable temperature insert operating in the temperature range \qtyrange[range-units = single]{2}{300}{K}. The external magnetic field was applied by a superconducting solenoid. The setup is based on a mechanical resonator driven by a voice coil actuator (VCA). A commercially available lock-in amplifier (Stanford Research Systems, SR 830) provides the
reference signal for the VCA, which is amplified by an audio amplifier with an output power of up to \qty{100}{W}. The magnetic sample signal was picked up by two coils arranged as first-order gradiometer connected to the lock-in amplifier signal input. A diamagnetic drinking straw with an inner diameter of 4\,mm was used as sample holder for in-plane and out-of-plane measurements. Reference measurements were done with a permalloy thin film on sapphire in both geometries for signal calibration. The oscillation amplitude was continuously measured using a quadrature
encoder. This data was used to correct for possible amplitude variations for every data point. The obtained absolute magnetic moments versus magnetic field where corrected with regard to the diamagnetic background of substrate and sample holder.
%
%
\subsection{Micromagnetic simulations}
Micromagnetic simulations have been performed using \textit{OOMMF} \cite{Donahue1999} as computational backend or micromagnetic solver in conjunction with \textit{Ubermag} as frontend, a collection of several Python packages \cite{Beg2022}. The gyromagnetic ratio $\gamma_0 = \mu_0\gamma$ and damping constant $\alpha$ in the dynamic equation (Landau-Lifshitz--Gilbert equation) were set to \qty{2.211e5}{m/As} and 0.1, respectively. The \textit{Ubermag} class \textit{HysteresisDriver} was used for magnetic hysteresis simulation \cite{Beg2022}. Exchange, Zeeman, cubic anisotropy and demagnetization (for out-of-plane field orientation) terms were taken into account for the free energy density $f$ according to
\begin{equation}
	\begin{split}
    f = &-A\vec{m}\cdot \nabla^2\vec{m} - \mu_0M_s\vec{m}\cdot\vec{H} \\
    	&- K\left[m_x^2m_y^2 + m_y^2m_z^2 + m_x^2m_z^2\right] \\
    	&- \frac{1}{2}\mu_0M_s\vec{m}\cdot\vec{H}_d
    \end{split}
\end{equation}
where $\vec{m}$ is the unit vector in direction of the magnetization, $A$ the exchange stiffness, $M_s$ the saturation magnetization, $\vec{H}$
the external magnetic field, $K$ the cubic anisotropy constant and $\vec{H}_d$ the demagnetizing field, respectively.
%
%
\subsection{Ab initio calculations}
All  calculations were performed within \textit{ab initio} density functional theory (DFT) \cite{Hohenberg1964,Kohn1965}. 
We used the Vienna ab initio simulations package VASP in version 6.3.0 \cite{Kresse1993,Kresse1996,Kresse1996_2}. 
The planewave basis set and projector augmented waves (PAW) \cite{Bloechl1994,Kresse1999} were employed as implemented in VASP.
The exchange-correlation contribution was treated using the PBE functional \cite{Perdew1996}. To deal with the localized nature of Eu 4$f$ orbitals, we introduced a Hubbard $U$ correction (DFT+U)  with $U = 5$~eV 
\cite{Song2023}. The energy cutoff was set to \qty{500}{eV}.
Each calculation for the magnetic anisotropy energy (MAE) consists of a three-step procedure \cite{Shuyang2021}: In the first step, starting from an already relaxed structure at $\varepsilon = \qty{0}{\%}$ strain, the structure was further relaxed for different strain values $\varepsilon = \qty{\pm 0.5}{\%} $. The strain was introduced by using the fcc primitive unit cell with a strain tensor

\begin{equation}\label{eq:straintensor}
\hat{\varepsilon} =
\begin{pmatrix}
    1 & \varepsilon             & \varepsilon             \\
    \varepsilon           & 1   & \varepsilon             \\
    \varepsilon           & \varepsilon             & 1 
\end{pmatrix}.
\end{equation}

Thereafter, the unit cell shape, as determined by $\varepsilon$, was kept fixed, while the volume and the internal coordinates were optimized. The sign of $\varepsilon$ determines the type of strain, where positive (negative) values represent an elongation (shortening) along the \hkl[111] direction, and, correspondingly, a compressive (tensile) strain within the \hkl(111)-plane. The relaxation was performed initially using a $8\times8\times8$ grid with a force convergence criterion lower than \qty{1e-3}{eV/\angstrom} and  including magnetism, but without spin-orbit-coupling (SOC). In a second step, the self-consistent-field (SCF) energies, charge density and wave functions were calculated using a much denser $12\times12\times12$ grid. In the final step we included spin-orbit coupling and continued to converge the structure for a denser $k$-mesh of $24\times24\times24$ for several ferromagnetic vectors.
By comparing their energies and electronic structures, we extracted information on the MAE according to \textit{ab initio} DFT. 
This procedure was repeated for each value of strain. 

%
%
\section{Results and Discussion}
%
%
\subsection{Modeling the multi-domain morphology}
\begin{figure}[ht]
\centering
\includegraphics[width=\linewidth]{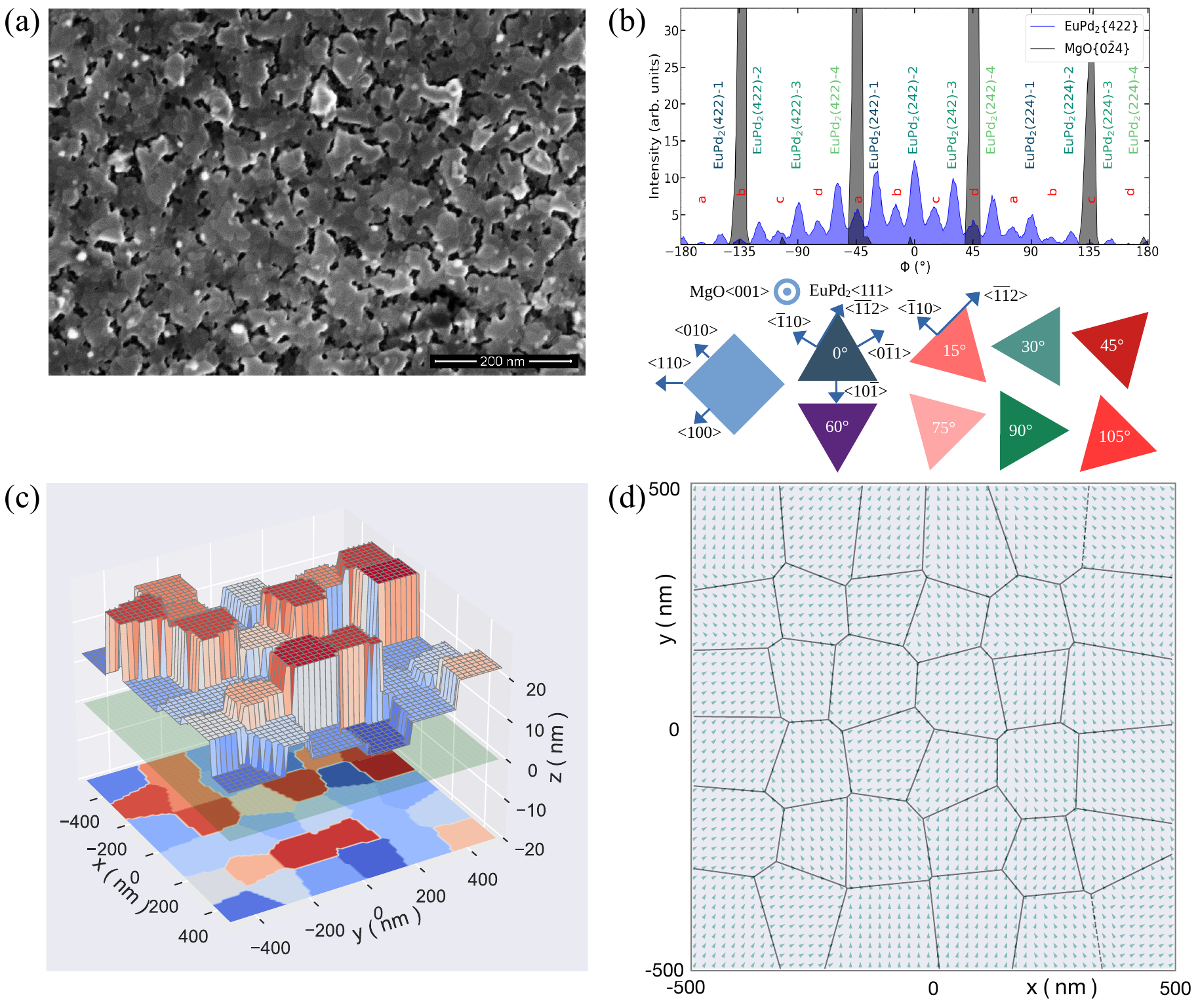}
\caption{(a) Scanning electron microscopy image of \ce{EuPd2}\hkl(111) multi-domain island growth. (b) Non-symmetric XRD pattern showing the multi-domain character as schematically indicated by the collection of triangles representing the different domain orientations (view from top). (c) Perspective view onto the multi-domain island growth morphology as used for the micromagnetic simulations; see text for details. (d) Top view of the multi-domain island structure indicating the respective local $\hat{x}$-axes orientation as projected into the film plane; see text for details.}
\label{fig2}
\end{figure}

As already alluded to in the methods section, the growth mode of the \ce{EuPd2} thin films is multi-domain with eight possible in-plane orientations for each of the islands; see Fig.\,\ref{fig2}(b). This is also apparent from scanning electron microscopy images taken \textit{ex situ} after growth where the island morphology with well-defined island edges is clearly visible, as is shown in Fig.\,\ref{fig2}(a).

As will be discussed below, the low-temperature magnetic hysteresis behavior of the \ce{EuPd2}\hkl(111) thin films is that of a soft ferromagnet with dominating shape anisotropy contribution for perpendicular magnetic field and a small coercive field for in-plane field orientation; see Fig.\,\ref{fig3}. However, in order to gain a more detailed understanding of the magnetic properties within a micromagnetic simulation approach, the multi-domain nature and morphology of the thin film has to be taken into account. To this end we apply the following approach: first, we define a square area $w\times w=1000\times 1000$\,nm in the xy-plane and divide it into regularly spaced square elements with the side $w_i = w/6$ and height $h = \qty{20}{nm}$. The length $w_i$ is chosen so as to roughly represent a typical diameter of the growth islands of the \ce{EuPd2} thin film, whereas the height is chosen to represent the average island height (as deduced from atomic force microscopy measurements; not shown). Next, we randomly offset the centers of the square elements in x and y choosing from a uniform distribution in the limits $[0, \mathrm{max})$ with $\mathrm{max} = w_i/4$. This is followed by the generation of a kd-tree from the collection of element centers. A kd-tree is a binary tree for point-neighborhood classification based on a hierarchical subdivision of space by splitting hyperplanes that are orthogonal to the coordinate axes. We use the class kd-tree from the Python package Scipy 1.11.3 for this and its associated query function \cite{Scipy2020}. We also randomly change the height of the elements by choosing from a uniform distribution in the limits $(-dh, +dh)$ with $dh = \qty{10}{nm}$ in order to model the different heights of the EuPd$_2$ growth islands.

Using the kd-tree structure we can find the nearest of the center points for any given point. Using this information the model thin film morphology is now fully defined and is shown in Fig.\,\ref{fig2}(c). The kd-tree structure is also used in the micromagnetic simulations in which for any given point in the simulation volume querying the kd-tree allows to decide which island this point belongs to or if this point is just in empty space.

In order to take the 8 different in-plane orientations of the islands into account, we select for each island randomly any of the 8 domain orientations and define for each island the corresponding directions of the local x- and y-axis. In Fig.\,\ref{fig2}(d) we show the result for our model thin film morphology. The small arrows indicate the respective local direction of the positive $\hat{x}=\hkl[100]$-axis as projected into the plane perpendicular to the \hkl[111] growth direction.
%
%
\subsection{Pre-analysis of magnetic hysteresis behavior}
\begin{figure}[ht]
\centering
\includegraphics[width=.95\linewidth]{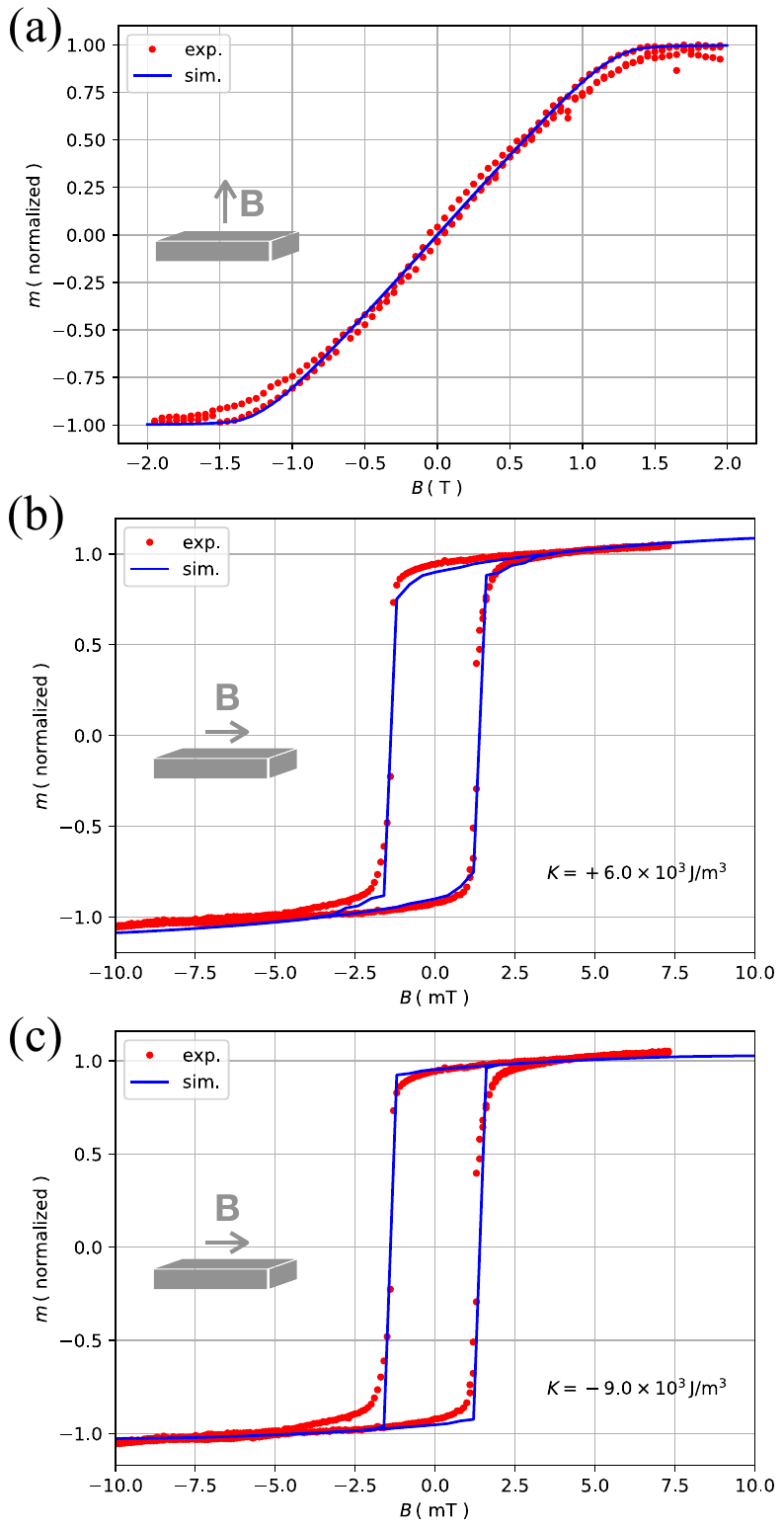}
\caption{(a) Normalized magnetization $m$ vs. magnetic field $B$ for field perpendicular to the film plane at $T = \qty{5}{K}$. The red dots are experimental data taken with a vibrating sample magnetometer. The blue curve shows the result of the micromagnetic simulation. (b)/(c) Experimental magnetization data vs. magnetic field for field in the thin film plane in red at $T = \qty{5}{K}$. The blue lines represent the results of the micromagnetic simulations assuming cubic crystal anisotropy and periodic boundary conditions using the anisotropy constant $K$ as indicated.}
\label{fig3}
\end{figure}

In Fig.\,\ref{fig3}(a) and (b)/(c) the low-temperature magnetization vs.\ external field dependence of the \ce{EuPd2}\hkl(111) thin film is shown for two different field orientations, as indicated. The large field needed to saturate the magnetization for perpendicular field and the small coercive field observed for in-plane field orientation indicate the overall behavior of a soft ferromagnet. This means that crystal anisotropy contributions are negligible for perpendicular field and the saturation magnetization $M_s$ can directly be obtained from the observed saturation field $B_s = \mu_0 M_s$. We find $M_s = \qty{1.09e6}{A/m}$ (i.\,e.\ $6.89\,\mu_B$/Eu(II)) which corresponds to \qty{92}{\%} of the value found for single crystals \cite{Nakamura2016}. This value for the saturation magnetization at low temperatures is used in the following micromagnetic simulations.
%
%
\subsection{Micromagnetic simulations}
Besides $M_s$, the exchange stiffness $A$ and the crystal anisotropy constant $K$ for cubic crystal symmetry have to be specified for the micromagnetic simulations. $K$ can be deduced from the observed coercive field for in-plane field orientation (see below). Regarding the exchange stiffness we proceeded as follows: Using single crystal $M(T)$ data from Nakamura et al.\ \cite{Nakamura2016} we employed a recent \textit{ad hoc} approach by Kuzmin et al.\ \cite{Kuzmin2020}, who introduced a ``shape factor'' $s$ relating the spin-wave, $D$, and exchange, $A$, stiffness constants of ferromagnetic compounds to their spontaneous magnetization $M_s$ and the Curie temperature $T_C$, using the following relation:
\begin{equation}
D = 0.1509\left(\frac{g\mu_B}{s\beta M_0}\right)^{2/3}k_BT_C
\end{equation}
with $M_0 = M_s(T=0).$ 
They further showed that the temperature dependence of $M_s$ can be extremely well described by a semi-empirical formula
\begin{equation}
M_s(T) = M_0\left[ 1 - s\left( \frac{T}{T_C} \right)^{3/2} - (1 - s)\left( \frac{T}{T_C} \right)^p \right]^\beta
\label{eq:kuzmin_fit}
\end{equation}
where $p\geq 3/2$ and $s>0$ are adjustable parameters and $\beta$ is the critical exponent for the magnetic order parameter. Eq.\,\ref{eq:kuzmin_fit} was constructed to obey Bloch's power law at low temperatures and to reproduce the critical behavior as $T$ approaches $T_C$ \cite{Kuzmin2005}.

We found that Eq.\,\ref{eq:kuzmin_fit} describes the $M(T)$-data from Nakamura et al.\ exceedingly well, with $p=5.2$, $\beta = 1/4$ and $s=1.4$. $T_C$ and $M_0$ were set to \qty{82.0}{K} and \qty{1.19e6}{A/m}, respectively. Using the following relationship \cite{Kuzmin2020}
\begin{equation}
A = \frac{M_0}{2g\mu_B}D,
\end{equation}
which absorbs all details of the Heisenberg exchange interactions into one affective parameter $A$,
we obtain $A=\qty{6.9e-13}{J/m}$. which we used for the micromagnetic simulations. An accurate determination of the full Haisenberg Hamiltonian is difficult, but the results of the micromagnetic simulations with focus on magnetic hysteresis behavior do not sensitively depend on the exact value for $A$, as the exchange energy contribution is by far dominating, which lends credence to our approach. 

From the now known values for $M_s$ and $A$ we obtain the exchange length $via$ the relation $\ell_{ex} = \sqrt{A/(1/2\mu_0 M_s^2)}$ \cite{Rave1998}, which amounts to about 1\,nm. This small exchange length has to be considered in connection with the selection of the required minimum mesh size for the micromagnetic simulations. We tested the results of our simulations with mesh sizes of 1, 2 and 5\,nm for a smaller simulation volume of size $500\times 500\times (h+dh)$ and obtained virtually identical results. This is indeed expected as the crystal anisotropy -- indicated by the small coercive field for in-plane field orientation -- is small and for soft magnetic materials the rather smooth magnetization transitions can be resolved with minimum mesh sizes larger than the exchange length. We therefore used \qty{5}{nm} mesh size in all directions for the simulations with simulation volume $w\times w\times (h+dh)$.

Based on the thin film model morphology and local crystal axes information for each island, the micromagnetic simulations can now be performed. For each mesh element the information is used to provide the associated value for the saturation magnetization $M_s$ and directions of the local $\hat{x}=\hkl[100]$- and $\hat{y}=\hkl[010]$-axis. The overall simulation volume has dimensions $w \times w \times (h+dh)$. For in-plane field orientation we assume 2D periodic boundary conditions in order to avoid unphysical edge effects caused by the finite edge length $w$. In this geometry we neglect demagnetization contributions as the demagnetization factor is close to 0 for thin films in this field orientation. For field perpendicular to the film plane we take demagnetization effects into account but do not use periodic boundary conditions.

In Fig.\,\ref{fig3}(a)-(c) we show the result of the micromagnetic simulations as blue lines for both field orientations, as indicated. Excellent correspondence between simulation result and experimental data is observed for the perpendicular field orientation. Also, for in-plane field very good correspondence can be obtained assuming a cubic crystal anisotropy constant $K$ of $6\dots 9\times 10^3$\,J/m$^3$ (i.\,e.\ $2.3\dots 3.4\,\mu$eV/f.u.) with the exact value depending on the assumed easy axis direction; see Fig.\,\ref{fig3}(b,c). Due to the multi growth domain morphology it is not possible to unequivocally identify if either the $K > 0$ or $K < 0$ easy-axes behavior is a better fit to the date. The experimentally observed residual slope of the magnetization at higher fields does however indicate that $K>0$ easy axes behavior is more likely.

In the next subsection we address the question how the deduced magnetic anisotropy as reflected by $K$ can be reconciled with a presumably pure spin moment for \ce{Eu^{2+}} by taking into account that under cooling the thin film on \ce{MgO} to cryogenic temperature, uniform biaxial in-plane tensile strain in the range of 0.1\,\% develops due to clamping and differing thermal expansion coefficients for the Eu-compound and the substrate material. This was shown experimentally for \ce{EuPd2Si2} thin films on \ce{MgO} for which single crystal thermal expansion data as a function of temperature are available \cite{Koelsch2022}. For \ce{EuPd2} thin films on \ce{MgO} an analogous behavior is expected.
%
%
\subsection{Ab initio calculations and the consequences of 
strain}
It is well known that MAE in cubic crystals only appears in the fourth order in spin-orbit coupling (SOC), and is therefore very small, usually on the order of \unit{\mu eV}. However, upon lattice distortion the cubic constraints are linearly relieved and a second order anisotropy, in our case between $\langle 111\rangle$ and $\langle 1\bar{1}0\rangle$ reappears.

\begin{figure}[ht]
\centering
\includegraphics[width=\linewidth]{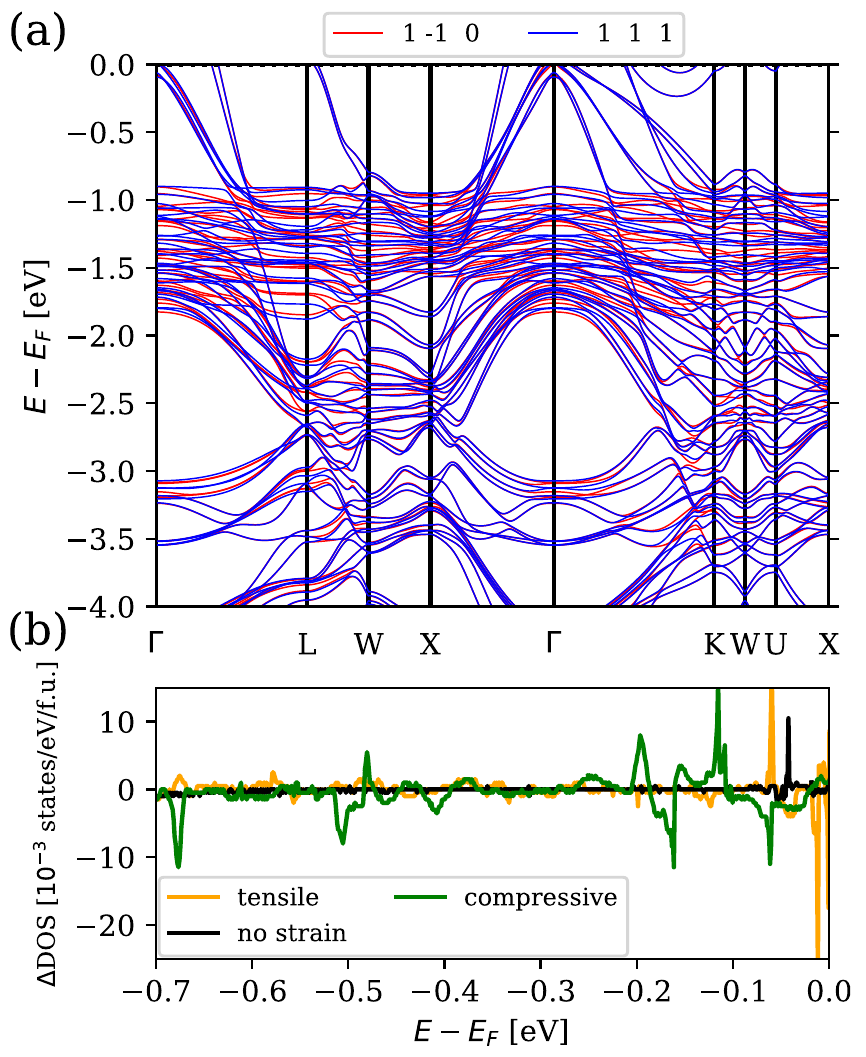}
\caption{Ab initio PBE+SOC+U($=5$~eV) band structure along an fcc high symmetry path for the cubic symmetry $(\varepsilon=0$\% for (a) magnetization in \EuPdII{} along \hkl(111) (blue) and \hkl(1-10) (red) directions. (b) Comparison of the difference of the density of states ($\Delta$DOS) for tensile (orange), compressive (green) $\varepsilon=\pm0.5\%$ and 0) and no strain (black).}
\label{fig:abinitio}
\end{figure}

In order to illustrate this numerically, we have calculated electronic bands and the total DFT energy with and without uniaxial strain along [111], as described above.  In Fig.\,\ref{fig:abinitio}(a) we show the band structure for undistorted FM \EuPdII{} for the magnetization oriented along \hkl[111] (blue) and \hkl[1-10] (red) directions. 
While some band crossings are removed linearly in SOC even in the undistorted case, because of the cubic symmetry they cancel out and, as a results, the anisotropy is so small that it is below the accuracy of our calculations. Also the density of states (Fig.\,\ref{fig:abinitio}(c)) is nearly identical for the two orientations. 
As mentioned, this cancellation is incomplete under strain (Fig.\, \ref{fig:abinitio}(b)). One may think that because the SOC effects in the band structure are most pronounced well below $E_F$, they should not affect the total energy, as the states below and above an avoided crossing cancel each other. It was shown long ago in seminal papers by Jansen \cite{Jansen1988,Jansen1999}, that while a crossing situated at an energy $E$ below the Fermi level indeed contributes only $\sim\xi^2/E$ to the band anisotropy, upon integrating over occupied states the contribution becomes, by order of magnitude, $\xi^2 N \log( W/\xi)$, where $N$ is the average density of states and $W$ the band width, thus acquiring a large logarithmic factor. In this way, Jansen observed, fully occupied states also contribute to MAE. This is a pure band effect and does not exist in free ions.

In Fig.\,\ref{fig:abinitio} we show the band structure for magnetization \EuPdII{} along \hkl<111> (blue) and \hkl<1-10> (red) directions. In the panel (a) we show the bands for the undistorted structure, and (b) for that compressed along \hkl<111> by $\varepsilon=+0.5\%$. In  Fig.\,\ref{fig:abinitio}(c) we compare densities of states for $\varepsilon=\pm0.5\%$ and 0.  These strains are much larger than those in the experiment, but for the smaller distortions the calculated anisotropy was to small to be reliably estimated.

In the following we discuss the significance of the results for the experimental observations. The MAE differences $E_{MAE}(\varepsilon=-0.5\%)=E_{1\bar{1}0}-E_{111}=-13\mu$eV/f.u. for tensile strain and $ E_{MAE}(\varepsilon=+0.5\%)=+13\mu$eV/f.u. are just indicative of the expected anisotropy constant used in the micromagnetic simulation. When converting to the same units as the anisotropy parameter of the experiment, we obtain $|K_{DFT}|=|E_{MAE}(\varepsilon=\pm0.5\%)|=13\mu$eV/f.u. $/V_{\varepsilon=\pm0.5\%}\simeq0.2\mu$  eV/\AA{}$^3\simeq 34.1 \times 10^3$ J/m$^3$. Taking the used strain values of 0.5\,\% into account and considering the actual expected strain of 0.1\,\%, we linearly extrapolate to $|K_{DFT}| \approx 6\times 10^3$\,J/m$^3$ which is quite similar to the result for $K$ from the micromagnetic simulation.


%
%
\section{Conclusion}
By combination of micromagnetic simulations and \textit{ab initio} calculations the importance of strain effects in understanding magnetic anisotropy effects of the pure spin ferromagnetism of the cubic Laves phase compound \ce{EuPd2} was studied for thin films. Despite the fact that both Pd and Eu in this compound formally have  closed magnetic shells ($d$- and $f$-, respectively), and are not supposed to have orbital moments, we can unambiguously detect magnetic anisotropy both in the experiment and in the calculations, and identify its microscopic origin. 

By taking the multi growth island domain structure of the films into account, quantitative agreement between the low-temperature magnetic hysteresis behavior with the micromagnetic simulation results for two different field orientations is obtained. 
%
%
\section{Acknowledgements}
The authors are grateful for funding by the Deutsche Forschungsgemeinschaft (DFG, German Research Foundation) Grant No.\ TRR 288 - 422213477 (Projects No.\ A04 and A05). IIM acknowledges support from the Office of Naval Research under the grant \#N00014-23-1-2480. He is also grateful to the Wilhelm
und Else Heraeus Stiftung for supporting his travel to
Germany.

\clearpage
\bibliography{EuPd2}

\end{document}